\documentclass[aps,prl,twocolumn,showpacs,superscriptaddress,longbibliography]{revtex4-2}
\usepackage{amsfonts}
\usepackage{amsmath}
\usepackage{graphicx}
\usepackage{bm}
\usepackage{amssymb}
\usepackage{dcolumn}
\usepackage{color}
\usepackage{soul}
\usepackage{siunitx}
\usepackage{multirow}
\usepackage{booktabs}
\usepackage{diagbox}
\usepackage{verbatim}

\usepackage[colorlinks,
    linkcolor=blue,
    anchorcolor=blue,
    citecolor=blue,
    urlcolor=blue,
]{hyperref}

\usepackage{fontspec}

\begin{document}

    \title{Out-of-Plane Oscillating Electric Fields Unlock Low-Temperature Nonequilibrium Superionicity in Quasi-Two-Dimensional AgCrSe$_2$}

    \author{Jia-Wen Li}
    \affiliation{Laboratory of Theoretical and Computational Nanoscience, National Center for Nanoscience and Technology, Chinese Academy of Sciences, Beijing 100190, China}
    \affiliation{University of Chinese Academy of Sciences, Beijing 100049, China}

    \author{Kun Yang}
    \affiliation{Laboratory of Theoretical and Computational Nanoscience, National Center for Nanoscience and Technology, Chinese Academy of Sciences, Beijing 100190, China}
    \affiliation{University of Chinese Academy of Sciences, Beijing 100049, China}

    

    \author{Sheng Meng}
    \affiliation{Beijing National Laboratory for Condensed Matter Physics, and Institute of Physics, Chinese Academy of Sciences, Beijing 100190, China}
    \affiliation{University of Chinese Academy of Sciences, Beijing 100049, China}
    \affiliation{Songshan Lake Materials Laboratory, Dongguan, Guangdong 523808, China}
    
    \author{Xinghua Shi}
    \affiliation{Laboratory of Theoretical and Computational Nanoscience, National Center for Nanoscience and Technology, Chinese Academy of Sciences, Beijing 100190, China}
    \affiliation{University of Chinese Academy of Sciences, Beijing 100049, China}

    \author{Wei-Hai Fang}
    \affiliation{College of Chemistry, Key Laboratory of Theoretical and Computational Photochemistry of Ministry of Education, Beijing Normal University, Beijing 100875, China}

	\author{Jin Zhang}
	\email{jinzhang@nanoctr.cn}
	\affiliation{Laboratory of Theoretical and Computational Nanoscience, National Center for Nanoscience and Technology, Chinese Academy of Sciences, Beijing 100190, China}
	\affiliation{University of Chinese Academy of Sciences, Beijing 100049, China}
	
\begin{abstract}   	
   	Superionic conductors enable exceptionally fast ion transport in solids and offer broad technological potential.
   	However, their highly conductive states are typically accessed only above the order--disorder transition of the mobile-ion sublattice, which limits low-temperature operation.
   	Here, electric-field-responsive machine-learning molecular dynamics reveals a nonequilibrium route to superionic transport in quasi-two-dimensional AgCrSe$_2$ driven by an out-of-plane oscillating electric field.
   	At 300 K, far below the transition temperature near 475 K, the field activates fast in-plane Ag$^+$ transport with an ionic conductivity reaching $\sim$1.7 S cm$^{-1}$.
   	This originates from field-driven dynamic disorder of Ag$^+$ across two equivalent sublattices rather than thermally induced disordering.
   	With increasing frequency, the Ag occupation evolves from complete switching to a highly conductive dynamically disordered state and finally to an ordered state with weak field response.
   	These regimes shift to higher frequencies with increasing temperature or field amplitude, consistent with competition between the driving period and Ag-sublattice response time.
   	At strong fields, this timescale is consistent with field-assisted thermal activation, while weak-field deviations suggest additional dynamics beyond single-ion activation.
 	These results suggest timescale-matched periodic driving as a route to transport-active ionic disorder far below an equilibrium superionic transition.
\end{abstract}
    \maketitle


	\textcolor{blue}{{\em Introduction}}---
	Superionic conductors (SICs) are characterized by exceptionally fast ion transport within a solid host, exhibiting ionic diffusivities and conductivities that can approach those of liquid electrolytes \cite{Jun2024,Sau2024}. 
	Upon crossing a characteristic order--disorder transition temperature $T_{\rm od}$, the mobile-ion sublattice becomes dynamically disordered, enabling rapid liquid-like migration through the solid. 
	Their exceptionally high ionic conductivity makes them promising candidates for various technological applications such as fuel cells and electrolysis cells \cite{Liu2023,Zhang2023c}, supercapacitors \cite{Zhi2023}, intercalation MXenes \cite{Lukatskaya2013}, electrolytes in lithium-ion \cite{Yin2023} and other solid-state batteries \cite{Kato2016,Boolchand2001}.

	One outstanding limitation to the widespread application of many SICs is that their highly conductive superionic phases are stable only at elevated temperatures \cite{Gu2024,Feng2022}.
	For example, the $T_{\rm od}$ of AgI and AgCrSe$_2$ are respectively 420 K and 475 K, substantially exceeding room temperature 300 K \cite{Hajibabaei2025,Ding2020}.
	Many efforts have been devoted to overcoming this limitation to promote fast ionic transport at lower temperatures, including size control of $\alpha$-AgI particles \cite{Makiura2009} and thickness reduction of AgCrS$_2$ \cite{Peng2021}, compressive strain in Cu$_2$Se and Li$_2$Se \cite{DumettTorres2018}, defect and vacancy engineering in LaH$_x$ \cite{Zhang2023c}, Ag$_2$Te \cite{Xu2026}, LiB$_3$ \cite{Yang2026a}, and AgCrSe$_2$ \cite{Wang2021}, as well as chemical substitution in KAg$_3$Se$_2$ \cite{Rettie2021}, Li$_3$YCl$_6$-type halides \cite{Liu2024}, Li$_2$ZrCl$_6$ \cite{Guan2026}, and CuCrSe$_2$ \cite{Rahman2025}.

	In contrast, electric fields can couple directly to ions, offering an effective route for manipulating ionic motion without modifying the material composition.	
	Terahertz (THz) electric fields have been used to directly accelerate ionic hopping in $\beta$-aluminas \cite{Minami2020,Poletayev2024}, while selective excitation of TiO$_6$ rocking modes can substantially enhance Li$^+$ hopping in Li$_{0.5}$La$_{0.5}$TiO$_3$ \cite{Pham2026}.
	Conversely, ultrafast optical excitation can drive AgCrSe$_2$ into the superionic state by rapidly heating the system above its $T_{\rm od}$, thereby activating fast Ag$^+$ migration \cite{Yang2026}.
	However, whether oscillating electric fields can induce superionic transport below $T_{\rm od}$ remains largely unexplored.

	In this work, we investigated the field-driven ionic dynamics of quasi-two-dimensional AgCrSe$_2$ using an electric-field-responsive machine-learning interatomic potential.
	An out-of-plane oscillating electric field is found to activate fast in-plane Ag$^+$ transport at 300 K, far below its equilibrium transition temperature $T_{\rm od}$ of 475 K, with a high ionic conductivity reaching 1.7 S cm$^{-1}$.	
	This originates from field-driven Ag$^+$ redistribution between two equivalent sublattices, producing the dynamic occupation disorder characteristic of the high-temperature superionic phase and enabling fast ionic migration.
	Correspondingly, the ionic conductivity shows a pronounced positive correlation with the dynamic disorder of the Ag sublattices.
	As the frequency increases, the Ag occupation across the two sublattices evolves from complete switching, through a highly conductive dynamically disordered state, to a largely ordered state showing weak response.
	Increasing temperature or field amplitude shifts this dynamic picture toward higher frequencies.
	These behaviors reflect a competition between the driving period and the characteristic time for Ag ions to cross the migration barrier between two lattices.
	At strong fields, this timescale follows field-assisted thermal activation, whereas the additional slowdown at	weak fields suggests dynamics beyond single-ion activation, such as the correlated Ag rearrangement.
	Together, these results reveal oscillating electric fields as a route to access low-temperature superionic transport, opening new possibilities for dynamically controlling superionic states.

	

\begin{figure}[hbpt]
		\centering
		\includegraphics[width=1.0\columnwidth]{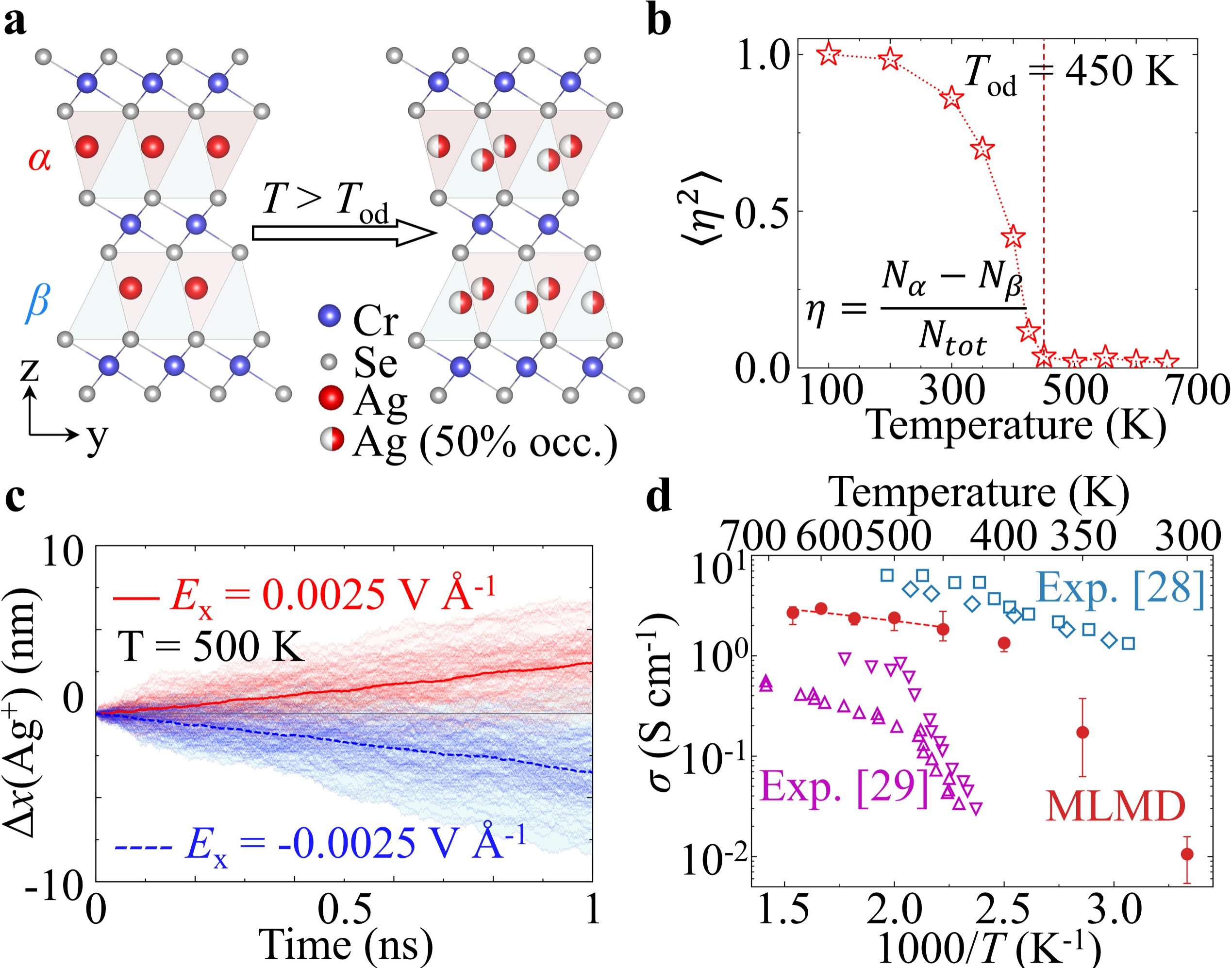}\\
		\caption{		
			(\textbf{a}) Crystal structure of AgCrSe$_2$ and its order--disorder transition.
			At low temperature, Ag$^+$ ions predominantly occupy the $\alpha$ sublattice, whereas in the high-temperature superionic state they dynamically populate the symmetry-equivalent $\alpha$ and $\beta$ sublattices with nearly equal probability.
			(\textbf{b}) Temperature dependence of $\langle \eta^2 \rangle$, where $\eta=(N_{\alpha}-N_{\beta})/N_{\mathrm{tot}}$ and $N_\alpha$ ($N_\beta$) is the number of Ag ions occupying the $\alpha$ ($\beta$) sites.
			The collapse of $\langle \eta^2 \rangle$ near 450 K marks $T_{\mathrm{od}}\simeq450$ K, close to the experimental value of 475 K \cite{Yang2026}.
			(\textbf{c}) Machine-learning molecular dynamics (MLMD) trajectories of Ag$^+$ at 500 K under $E_x=\pm0.0025$ V $\mathrm{\AA}^{-1}$.
			Despite strong thermal fluctuations in individual trajectories, the ensemble-averaged displacement reverses with the applied field.
			(\textbf{d}) Temperature-dependent ionic conductivity from MLMD.
			Error bars span the simulated $\sigma$ range at each temperature.
			Above 450 K, the conductivity follows Arrhenius behavior with an activation energy of 97 meV, whereas transport is strongly suppressed below 400 K.
			The calculated conductivities are consistent in magnitude and temperature dependence with experimental measurements \cite{Murphy1977,Boukamp1983}.
		}
		\label{Fig1}
	\end{figure}

	\textcolor{blue}{{\em Order-disorder transition and ionic transport in AgCrSe$_2$}}---
	The crystal structure of AgCrSe$_2$ is shown in Figure~\ref{Fig1}(a), where Ag$^+$ ions are intercalated between the CrSe$_2$ layers.
	At low temperatures, Ag$^+$ ions predominantly occupy the $\alpha$ sites of the two symmetry-equivalent sublattices $\alpha$ and $\beta$ (space group $R3m$).
	Above $T_{\rm od}=475$ K, the Ag sublattice becomes dynamically disordered with Ag$^+$ ions distributed equally over both sublattices ($R\bar{3}m$), and high ionic mobility \cite{Murphy1977,Boukamp1983,Ding2020,Yang2026,Li2018a,Gascoin2011,Zhang2026}.

	\begin{figure*}[hbpt]
		\centering
		\includegraphics[width=1.95\columnwidth]{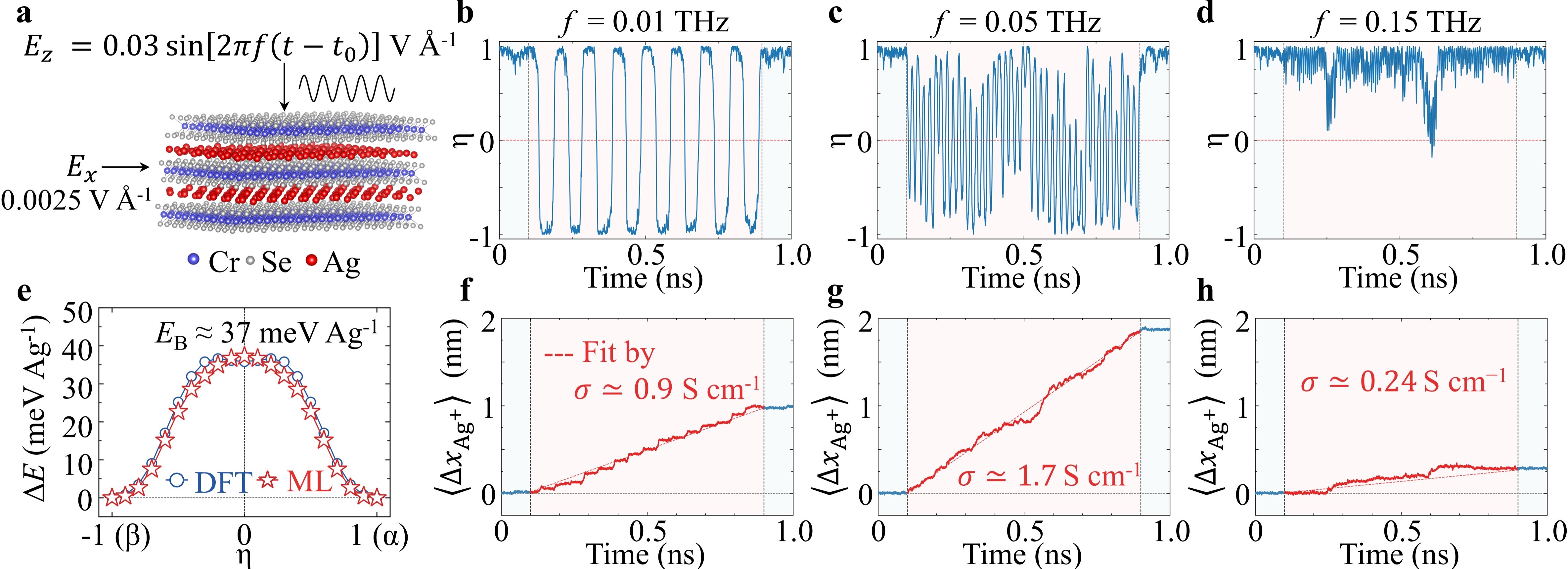}\\
		\caption{
			(\textbf{a}) Schematic of the electric-field-driven MLMD. 
			A sinusoidal out-of-plane field $E_z=0.03\sin[2\pi f(t-t_0)]$ V $\mathrm{\AA}^{-1}$ is applied, while an in-plane field $E_x$ of 0.0025 V $\mathrm{\AA}^{-1}$ probes Ag$^+$ transport.
			(\textbf{e}) The nudged elastic band energy profiles for the $\alpha$--$\beta$ transition from Density functional theory (DFT) and machine-learning (ML), showing consistent barriers of $\sim$ 37 meV Ag$^{-1}$.
			(\textbf{b, f}) At frequency $f=0.01$ THz, repeated switching of $\eta$ between +1 and -1 reflects complete $\alpha$--$\beta$ switching of the Ag sublattice.
			Ag$^+$ displacement increases during the switching intervals but nearly plateaus in the ordered $|\eta|\simeq1$ states, producing a stepwise displacement with $\sigma\simeq0.9$ S cm$^{-1}$.	
			(\textbf{c, g}) At $f=0.05$ THz, rapid incomplete $\alpha$--$\beta$ oscillations are observed, producing a dynamically disordered state with conductivity of 1.7 S cm$^{-1}$.
			(\textbf{d, h}) At $f=0.15$ THz, the Ag sublattice remains predominantly in the ordered state, with an ionic conductivity of 0.24 S cm$^{-1}$.
		}
		\label{Fig2}
	\end{figure*}

	To enable long-timescale simulations of Ag$^+$ dynamics, a machine-learning interatomic potential for AgCrSe$_2$ was trained using our electric-field-responsive graph neural network framework \cite{Li2026b}.
	Machine-learning molecular dynamics (MLMD) extends first-principles-level simulations to substantially larger length and time scales and has been widely applied to materials simulations \cite{Batatia2025,Wang2024a,Yu2025,Thiemann2026,Li2026b,Libbi2025,Ke2025,Li2026}.
	The dataset comprised 120 independent MD trajectories at 500--650 K in 50-K intervals, supplemented by low-temperature configurations at 300 and 400 K.
	The high-temperature superionic data were partitioned by trajectory to minimize correlations between subsets and prevent information leakage.
	The model shows high accuracy with mean absolute errors of 0.73 meV atom$^{-1}$ in energy and 33 meV \AA$^{-1}$ in force, with further details provided in the Supplementary Materials \cite{SM}.

	To track the temperature dependence of Ag-sublattice order, MLMD simulations were performed in a 432-atom supercell, as shown in Fig.~\ref{Fig1}(b).
	Each Ag ion was assigned to the $\alpha$ or $\beta$ sublattice according to the nearest reference site, with the order parameter defined as $\eta=(N_{\alpha}-N_{\beta})/N_{\rm tot}$, where $N_{\alpha}$ and $N_{\beta}$ are the corresponding Ag populations and $N_{\rm tot}=N_{\alpha}+N_{\beta}$.
	Thus, $\eta\simeq\pm1$ denotes an ordered state dominated by one sublattice, whereas $\eta\simeq0$ denotes comparable occupation of the two sublattices.
	As temperature increases, $\langle\eta^2\rangle$ decreases sharply and approaches zero near 450 K, yielding $T_{\rm od}\simeq450$ K, close to the experimental value of 475 K \cite{Murphy1977,Boukamp1983,Ding2020,Yang2026,Li2018a,Gascoin2011}.

	The ionic conductivity was subsequently evaluated by MLMD under in-plane electric fields.
	The field was incorporated through the additional force $\mathbf{F}_i^{\rm field}=\mathbf{Z}_i^*\mathbf{E}$, where $\mathbf{Z}_i^*$ is the configuration-dependent Born effective charge (BEC) tensor of atom $i$.
	The BECs were predicted with a neural network trained on the same configuration pool and partitioning protocol as the energy--force model, with metallic structures excluded based on the density of states at the Fermi level, yielding a mean absolute error of 0.10 $e$.
	The ionic conductivity was obtained as $\sigma_x=n_{\rm Ag}e v_x/E_x$, where $n_{\rm Ag}$ is the Ag number density, $e$ is the elementary charge, and $v_x$ is the mean Ag drift velocity induced by the in-plane probe field $E_x$.
	At 500 K, the ensemble-averaged Ag$^+$ displacement reverses with $E_x=\pm0.0025$ V \AA$^{-1}$ despite strong fluctuations in individual trajectories, as shown in Fig.~\ref{Fig1}(c).
	Conductivities were extracted from linear fits to the drift velocities at $E_x=\pm0.0010$ and $\pm0.0025$ V \AA$^{-1}$ and are shown in Fig.~\ref{Fig1}(d).
	Above the calculated $T_{\rm od}\sim450$ K, the ionic conductivity remains high and follows Arrhenius behavior with a 97-meV activation energy consistent with previous reports \cite{Wang2021,Brueesch1983,Murphy1977}, whereas it is strongly suppressed below 400 K.
	The calculated results provide reasonable estimates of the ionic conductivity and its temperature dependence compared with available experiments \cite{Murphy1977,Boukamp1983}.
	Temperature-dependent transport behavior and $\eta$ in the MLMD trajectories are provided in the SM \cite{SM}.


	\textcolor{blue}{{\em Out-of-plane oscillating-field-induced dynamic disorder and superionic transport}}---We next examined the nonequilibrium Ag$^+$ dynamics under an out-of-plane oscillating electric field.
	As schematically illustrated in Fig.~\ref{Fig2}(a), a sinusoidal field $E_z(t)=E_z^0\sin[2\pi f(t-t_0)]$ was applied, with $E_z^0$ and $f$ denoting the field amplitude and frequency, while an in-plane field $E_x=0.0025$ V \AA$^{-1}$ was used to probe Ag$^+$ transport.
	The associated $\alpha$--$\beta$ transition was further characterized by nudged elastic band calculations, as shown in Fig.~\ref{Fig2}(e), for which density functional theory and the machine-learning potential give consistent energy profiles and a barrier of approximately 37 meV per Ag.
	Under this driving protocol, MLMD simulations at 300 K and $E_z^0=0.03$ V \AA$^{-1}$ reveal three distinct frequency-dependent regimes: complete sublattice switching, a highly conductive dynamically disordered state, and, at higher frequencies, a largely ordered state with weak field response.
	Representative dynamics are shown in Fig.~\ref{Fig2}(b--d, f--h).

	At a low frequency of $f=0.01$ THz, the Ag sublattice undergoes repeated and complete $\alpha$--$\beta$ switching synchronized with the oscillating cycles, as shown in Fig.~\ref{Fig2}(b), reducing the time-averaged $\langle\eta^2\rangle$ from nearly 1 in the initial ordered state to $\sim0.76$.
	Correspondingly, the ensemble-averaged Ag$^+$ displacement along $E_x$ increases stepwise, with net drift occurring predominantly during the switching events and remaining negligible in the ordered $|\eta|\simeq1$ states.
	By contrast, Ag ions remain nearly immobile without the oscillating excitation.
	The resulting ionic conductivity reaches $\sim0.9$ S cm$^{-1}$, nearly two orders of magnitude higher than the equilibrium value at 300 K in Fig.~\ref{Fig1}(d).

	\begin{figure}[hbpt]
		\centering
		\includegraphics[width=1.0\columnwidth]{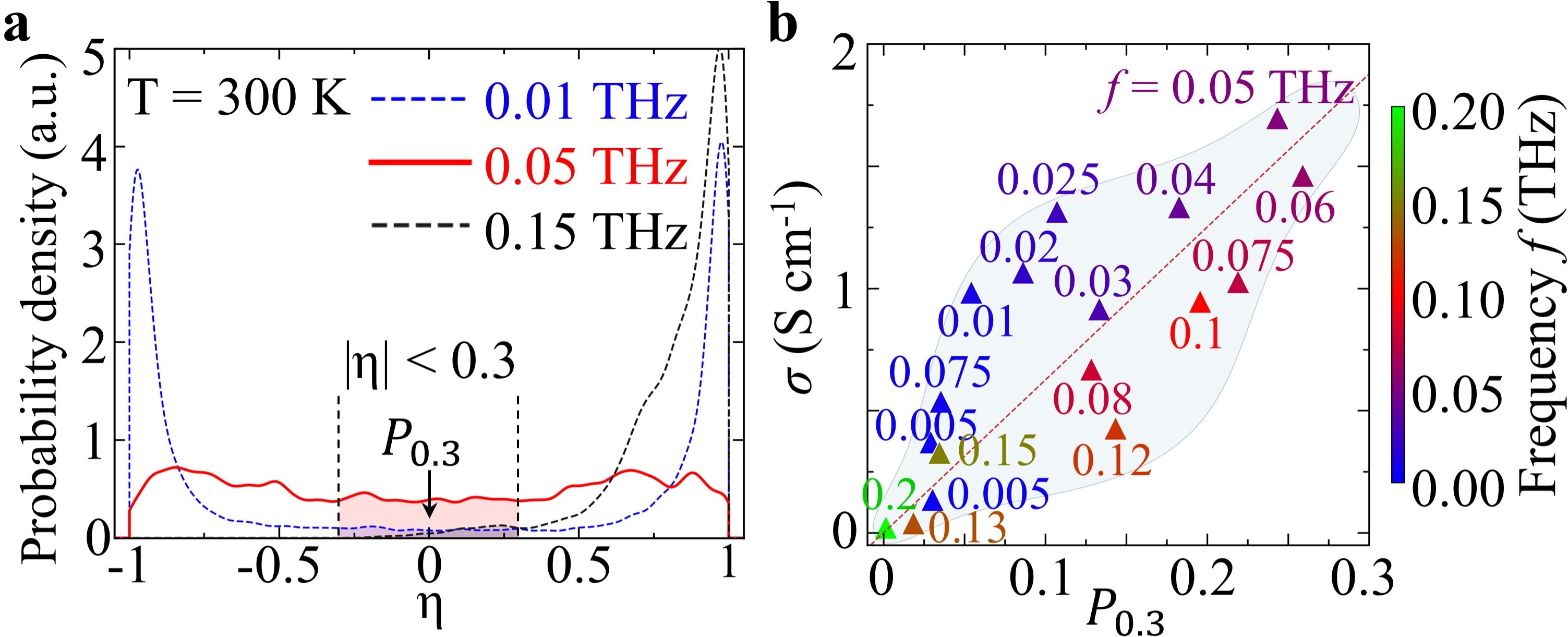}\\
		\caption{
			(\textbf{a}) Probability density of $\eta$ at 300 K under different driving frequencies.
			$P_{0.3}=P(|\eta|<0.3)$ quantifies the fraction of time spent in the dynamically disordered regime.
			(\textbf{b}) Distribution of $\sigma$ and $P_{0.3}$ across different driving frequencies, revealing a clear positive correlation.
		}
		\label{Fig3}
	\end{figure}
	
	At $f=0.05$ THz, the order parameter $\eta$ exhibits rapid oscillations without fully completing the $\alpha$--$\beta$ transition, producing pronounced dynamic disorder with $\langle\eta^2\rangle\simeq0.38$ and an ionic conductivity of approximately 1.7 S cm$^{-1}$, close to that of the equilibrium superionic state, as evident in Fig.~\ref{Fig2}(c, g).
	As the frequency increases to 0.15 THz, the Ag sublattice remains predominantly in the initial $\alpha$ state and only rarely becomes the dynamically disordered $\eta\simeq0$ region, reducing the ionic conductivity to approximately 0.24 S cm$^{-1}$, as evident in Fig.~\ref{Fig2}(d, h).
	Notably, the nanosecond-long simulations serve to ensure sufficient statistical sampling, rather than indicating a required nanosecond excitation duration.

	To quantify the connection between dynamic disorder and ionic conductivity, the probability distribution of the order parameter $\eta$ at different driving frequencies was analyzed.
	As shown in Fig.~\ref{Fig3}(a), the distribution is concentrated near $\eta\simeq\pm1$ at $f=0.01$ THz, becomes substantially broadened with significant weight around $\eta\simeq0$ at $f=0.05$ THz, and returns predominantly to the initial ordered state at $f=0.15$ THz.
	Accordingly, a probability $P_{0.3}=P(|\eta|<0.3)$ was introduced to quantify the residence in a highly disordered regime associated with superionic transport.
	Across the full frequency range from 0.005 THz to 0.200 THz, the ionic conductivity $\sigma$ shows a pronounced positive correlation with $P_{0.3}$, as shown in Fig.~\ref{Fig3}(b).
	This trend is consistent with previous simulations \cite{Wang2021}.
	The elemental temperatures remain centered near 300 K without sustained heating under the Langevin thermostat, as detailed in the SM \cite{SM}, indicating that the enhanced transport is not thermally induced.
	This correlation indicates that enhanced Ag$^+$ transport is directly associated with oscillating-field-induced dynamic disorder of the Ag sublattice.	
	Further transport analyses and $P_{0.3}$ statistics are provided in the SM \cite{SM}.


	\textcolor{blue}{{\em Field-driven ionic dynamics and characteristic switching timescale}}---To systematically map the field-driven transport response, MLMD simulations were performed across different driving field amplitudes, frequencies, and temperatures.
	Based on the strong correlation between ionic conductivity and dynamic Ag-sublattice disorder, $P_{0.3}$ was adopted as an operational criterion, with $P_{0.3}\geq10\%$ defining the transport-active regime and $P_{0.3}<10\%$ the transport-suppressed regime.
	The field--frequency maps at 250 and 300 K reveal a clear high-frequency cutoff above which transport is suppressed.
	The cutoff shifts to higher frequencies with increasing field amplitude and temperature, demonstrating that stronger driving and thermal fluctuations allow the transport-active disordered state to persist under progressively faster oscillations.
	The evolution of this cutoff points to a characteristic switching timescale.

\begin{figure}[hbpt]
		\centering
		\includegraphics[width=1\columnwidth]{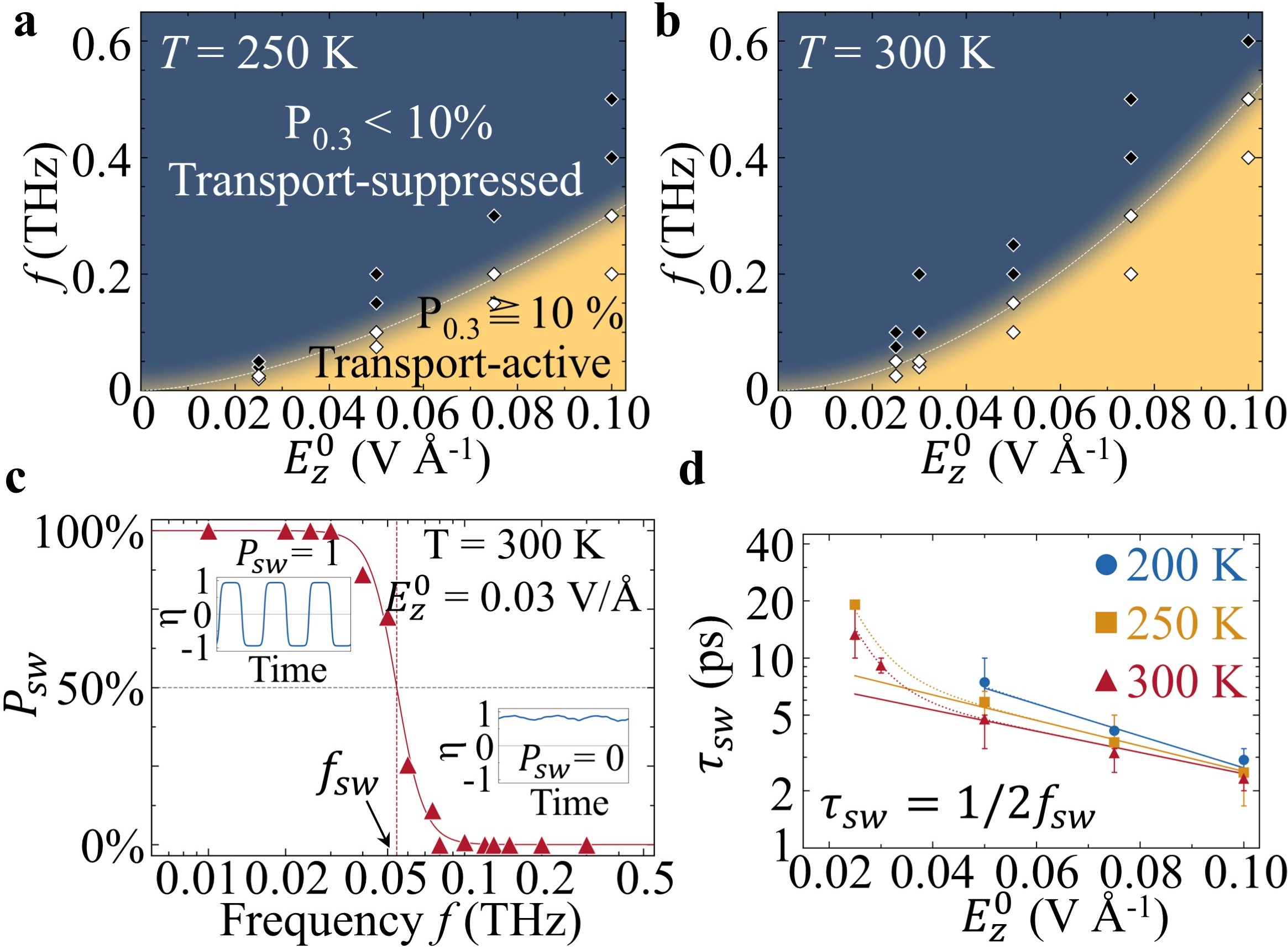}\\
		\caption{
			\textbf{(a,b)} Field--frequency diagrams at 250 and 300 K classified by $P_{0.3}=P(|\eta|<0.3)$.
			The transport-active regime is defined by $P_{0.3}\geq10\%$ (yellow), while $P_{0.3}<10\%$ denotes transport suppression (blue), with open and filled diamonds marking the corresponding simulated conditions, respectively.
			\textbf{(c)} Switching probability $P_{\rm sw}$ at 300 K and $E_z^0=0.03$ V \AA$^{-1}$.
			A complete switching event is counted when $\eta$ reaches both $+0.6$ and $-0.6$ within one field cycle, and $f_{\rm sw}$ is defined by $P_{\rm sw}=0.5$.
			Insets schematically illustrate complete-switching and non-switching cases.
			\textbf{(d)}  Characteristic switching time $\tau_{\rm sw}=1/(2f_{\rm sw})$ at 200, 250, and 300 K.
			Error bars indicate the nearest sampled points on either side of $P_{\rm sw}=0.5$.
			Solid curves denote the strong-field Arrhenius baseline, while dotted curves include a phenomenological correction for the additional weak-field slowdown.
		}
		\label{Fig4}
	\end{figure}

	To extract this timescale, complete $\alpha$--$\beta$ switching was quantified through the switching probability $P_{\mathrm{sw}}$, defined as the fraction of field periods in which $\eta$ reaches both +0.6 and -0.6.
	The result at 300 K and $E_z^0=0.03$ V \AA$^{-1}$ with different frequencies is shown in Fig.~\ref{Fig4}(c).
	$P_{\mathrm{sw}}$ is nearly 1 at 0.01 THz, where the 50-ps half-cycle provides sufficient time for complete switching.
	As the frequency increases from approximately 0.02 to 0.1 THz, $P_{\mathrm{sw}}$ decreases sharply from nearly 1 to almost 0, demonstrating that complete switching becomes progressively suppressed as the available half-cycle duration is shortened.
	At 0.1 THz, the half-cycle is reduced to only 5 ps, which is insufficient for nearly all complete switching events.
	The characteristic switching frequency $f_{\mathrm{sw}}$ is taken as the frequency at which $P_{\mathrm{sw}}=0.5$, and the corresponding half-period, $\tau_{\mathrm{sw}}=1/\left(2f_{\mathrm{sw}}\right)$ is taken as the characteristic switching time.
	For $T=300$ K and $E_z^0=0.03$ V \AA$^{-1}$, $f_{\mathrm{sw}}$ is approximately 0.055 THz, corresponding to $\tau_{\mathrm{sw}}\sim9$ ps.

	The characteristic switching times $\tau_{\rm sw}$ at 200, 250, and 300 K are shown as a function of $E_z^0$ in Fig.~\ref{Fig4}(d).
	A systematic acceleration of switching is observed upon either increasing the field amplitude or raising the temperature.
	In the strong-field regime, $\tau_{\rm sw}$ exhibits a near-linear dependence on $E_0$ on the semilogarithmic plot, suggesting an exponential field dependence of the switching time.
	Accordingly, the strong-field data were described using a field-assisted thermal-activation model  \cite{Kemp2024,Menzel2013,Salles2020},
	$
	\tau_A
	=
	\tau_0
	\exp\left[
	\frac{E_B-qE_z^0d_z}{k_BT}
	\right]
	$.
	Here, $\tau_A$ is the characteristic time predicted by the model, $\tau_0$ is the attempt-time prefactor, $E_B$ is the zero-field migration barrier, $q$ is the charge of an Ag$^+$ ion, $E_z^0$ is the oscillating-field amplitude, $d_z$ is the effective half-displacement of the Ag$^+$ ion projected along the field direction, $k_B$ is the Boltzmann constant, and $T$ is the temperature.
	With $E_B=37$ meV, $q=+e$, and $d_z=0.33$ \AA, the strong-field data ($E_z^0\geq0.05$ V \AA$^{-1}$) at all three temperatures are described using a single fitted prefactor $\tau_0=2.1$ ps.
	As shown by the solid lines in Fig.~\ref{Fig4}(d), the resulting baseline provides a good description of the strong-field switching times at all three temperatures, consistent with thermally activated crossing of a field-lowered migration barrier.
	Details of $P_{\rm sw}$ and $P_{0.3}$ are provided in SM \cite{SM}.

	At weaker fields with $E_z^0<0.05$ V \AA$^{-1}$, $\tau_{\rm sw}$ rises markedly above the Arrhenius baseline, suggesting dynamics beyond simple local barrier crossing.
	This may reflect the additional time required for correlated ionic rearrangement preceding the $\alpha$--$\beta$ transition.
	Correlated Ag motion has been observed previously \cite{Wang2021,Yang2026} and is also discovered in our MLMD trajectories (in SM \cite{SM}).
	In addition, the static calculations indicate eV-scale energetic penalties per transformed Ag for partial rearrangements involving either a single layer or a localized subset of Ag ions, far above the $\sim37$ meV per Ag barrier for collective $\alpha$--$\beta$ motion.
	At weaker fields, such correlated configurations may require longer times to develop, thereby delaying access to the low-barrier pathway.
	Therefore we adopted a phenomenological weak-field enhancement of $\tau_A$ by
	$
	\tau_{\rm sw}
	=
	\tau_A
	\left[
	1+\exp(u-\beta E_z^0)
	\right].
	$
	Here, $u$ is a dimensionless parameter controlling the magnitude of the correction, while $\beta$ determines its field dependence.
	The correction factor approaches unity at strong fields, recovering the field-assisted Arrhenius behavior, but increases rapidly as $E_z^0$ decreases, capturing the additional slowdown in the weak-field regime.
	With all Arrhenius parameters fixed at the values obtained from the strong-field analysis, only $u$ and $\beta$ were fitted globally to the data at all three temperatures, yielding $u=4.5$ and $\beta=173$.
	The resulting model describes the switching times over the full field range.

    \textcolor{blue}{{\em Conclusion}}---In conclusion, using MLMD, we reveal that out-of-plane oscillating electric fields can activate nonequilibrium superionic transport in quasi-two-dimensional AgCrSe$_2$ far below its equilibrium order--disorder transition temperature. 
    At 300 K, field-driven redistribution of Ag$^{+}$ between two equivalent sublattices produces a dynamically disordered state with an ionic conductivity reaching 1.7~$\mathrm{S\,cm^{-1}}$.
    The resulting dynamics reflect a competition between the driving period and the characteristic response time of the Ag sublattice.
    With increasing frequency, the Ag sublattice evolves from complete switching to a highly conductive dynamically disordered regime, and finally to an ordered state with weak field response. 
    The pronounced positive correlation between ionic conductivity and $P_{0.3}=P(|\eta|<0.3)$ connects enhanced transport with sustained access to the dynamically disordered regime.
    Increasing either field amplitude or temperature shifts the transport-active regime toward higher frequencies, reflecting a faster Ag-sublattice response.
    In the strong-field regime, the switching timescale follows a field-assisted thermal-activation picture, whereas the additional weak-field slowdown suggests dynamics beyond simple barrier crossing, possibly involving correlated ionic rearrangement.
    No sustained heating is observed, indicating that the enhanced transport does not originate from a temperature-driven transition.
    Together, these results suggest oscillating electric fields as a promising approach for accessing low-temperature superionic transport through timescale matching, and dynamically tuning ionic transport properties.


	\textcolor{blue}{{\em Acknowledgements}}---
	Jin Zhang acknowledges start-up support from the National Center for Nanoscience and Technology and funding from the National Natural Science Foundation of China (Grant No. 12574255).
	Jia-Wen Li acknowledges support from the Postdoctoral Fellowship Program (Grade C) of the China Postdoctoral Science Foundation (Grant No. GZC20261824).
	Xinghua Shi was supported by the National Key R\&D Program of China (Grant No. 2022YFA1203200) and the National Natural Science Foundation of China (Grant No. 12125202).
	Sheng Meng acknowledges support from the Chinese Academy of Sciences (Grant No. YSBR-047) and the National Natural Science Foundation of China (Grant No. 12450401).

    %

\end{document}